\documentclass[twocolumn,showpacs,preprintnumbers,amsmath,amssymb,floatfix]{revtex4-1}

\usepackage{multirow}
\usepackage{graphicx}
\usepackage{longtable}
\usepackage{bm}
\usepackage{dcolumn}
\usepackage{fancybox}

\newcommand{\GW}{$GW$}

\newcommand{\bfq}{{\bf q}}

\newcommand{\bfr}{{\bf r}}

\newcommand{\bfT}{{\bf T}}
\newcommand{\bfG}{{\bf G}}
\newcommand{\bfa}{{\bf a}}
\newcommand{\bfA}{{\bf A}}

\newcommand{\bfn}{{\bf n}}

\def\H0{H^0}

\newcommand{\req}[1]{\mbox{Eq.~\!(\ref{#1})}}

\def\connect#1{\leavevmode{\setbox1=\hbox{#1}\copy1%
\raise .2\ht1 \vbox{\moveleft \wd1\vbox{\hrule width \wd1 height .5pt depth 0pt}}%
}}

\def\ftn[#1]{\rlap{\footnotemark[#1]}}

\begin{document}

\title{Contracted Plane Wave satisfying periodic gauge}
\author{Takao Kotani}
\affiliation{Department of Applied  Mathematics and Physics, Tottori university, Tottori 680-8552, Japan}

\date{\today}
\begin{abstract}
We introduce the contracted plane waves (CPWs),
which satisfy the periodic gauge in the Brillouin-zone torus
as in the case of usual Wannier functions.
CPWs are very simply given as the sum of plane waves.
We will be able to use CPWs instead of the Wannier functions 
for the interpolation of physical quantities given
as the function of wave vectors in BZ.
Furthermore, it will be easy to complement the set of Gaussian bases by CPWs.
\end{abstract}
\maketitle

\section{Introduction}
\label{intro}
In the first-principles methods such as the pseudopotential (PP), LAPW, and
PAW, we use plane waves (PWs) or augmented plane waves (APWs)
for the bases to expand eigenfunctions \cite{rmartinbook}.
Even the linearized Muffin-tin orbital (MTO) method have recently evolved to be 
the APW plus MTO method (the PMT method) \cite{pmt1}, where
we can perform accurate and robust calculations
with the basis set of the highly-localized MTOs (without
material-dependent parameters)
complemented by the set of the low-energy-cutoff ($\lesssim$ 4 Ry) APWs \cite{kotani_formulation_2015,ecalj}. 
In contrast to these methods utilizing PWs/APWs, it is not so easy to obtain high-energy states accurately
in the methods with Gaussians as bases \cite{crystal},
because of the difficulty to enlarge the number of the Gaussian bases
systematically.
Hereafter, discussion is seemingly for PWs, however,
also for APWs with trivial modifications.

Although the set of PWs is very useful, the set is often
inconvenient for kinds of applications. Especially, PWs
are not suitable for the interpolation in the Brillouin zone (IinBZ) for variety of quantities.
IinBZ means evaluating matrix $F(\bfq)$ at any
$\bfq$ point in BZ from $\{F(\bfq)|\bfq \in {\rm mesh \ points  \ in \ the \ BZ}\}$.
As an example, let us consider the case to perform the $GW$ calculations
in the PW basis \cite{Hybertsen86,hamann_maximally_2009}.
Then we treat the non-local one-particle effective Hamiltonian
$H_\bfq(\bfG,\bfG')=\frac{1}{N}\int d^3r \int d^3r' \exp(-i (\bfq+\bfG) \bfr')
H(\bfr,\bfr') \exp(i (\bfq+\bfG') \bfr)$. 
Here $\bfG$ denotes the reciprocal vectors of a crystal; $N$ denotes
the number of primitive cells in the Born--von Karman boundary
condition. Because the \GW calculations are expensive,
we can calculate $H_\bfq(\bfG,\bfG')$
only at limited number of $\bfq$ in BZ.
Thus we need to obtain $H_\bfq(\bfG,\bfG')$ at any $\bfq$ via IinBZ.
IinBZ is very important, for example, to calculate effective mass
and/or the Fermi surfaces in the \GW calculations.
IinBZ is essential to perform stable QS\GW calculations 
\cite{kotani_quasiparticle_2014,kotani07a,esmqsgw2020}.
IinBZ is a key methodology even when we evaluate other kinds of quantities
such as electron-electron interaction, electron-phonon interactions,
topological quantities, and so on.

The difficulty of the IinBZ 
is due to the non-periodicity of the PWs $\exp(i (\bfq+\bfG)\bfr)$ with
respect to $\bfq$; when $\bfq$ changes across BZ to be $\bfq+\bfG_{\rm shift}$, 
$\exp(i (\bfq+\bfG)\bfr)$ changes to $\exp(i \bfG_{\rm shift}\bfr)\exp(i (\bfq+\bfG)\bfr)$.
That is, $\exp(i (\bfq+\bfG)\bfr)$ do not satisfy the periodic gauge
\cite{vanderbilt_2018}, $\exp(i (\bfq+\bfG)\bfr) \ne \exp(i \bfG_{\rm shift}\bfr)\exp(i (\bfq+\bfG)\bfr)$.
This ends up with $H_\bfq(\bfG,\bfG')\ne H_{\bfq+\bfG_{\rm shift}}(\bfG,\bfG')$.
This causes difficulty to use PWs for IinBZ.

To avoid this difficulty,
the so-called Wannier interpolation (WI) \cite{hamann_maximally_2009} is
introduced as a method for IinBZ.
After we construct a set of the Wannier functions by the procedure of
maximally localized Wannier function
\cite{marzari_maximally_1997,souza01,marzari_maximally_2012},
we re-expand $H_\bfq(\bfG,\bfG')$ by the Wannier functions
instead of PWs as $H_\bfq(i,i')$, where $i,i'$ are the the indexes of
the Wannier functions.
Since the Wannier functions satisfy the periodic gauge condition
\cite{vanderbilt_2018}, we can easily interpolate $H_\bfq(i,i')$ for any $\bfq$.
However, the construction of the maximally-localized
Wannier functions usually used in the WI are not so simple \cite{wang_first-principles_2019}.
We need to choose initial conditions and choose energy windows.
It is not so easy to make the method automatic without examination by human.
This causes a difficulty when we apply the method to the material's informatics
where we have to analyze thousands of possible materials in a work. 
In order to avoid the difficulty of WI, Wang et al. suggested
to use the first-principles method with the atomic-like localized
bases satisfying the periodic gauge \cite{wang_first-principles_2019}.
However, this suggestion do not remove a problem in WI;
for supercell with huge vacuum region,
we have to fill the region by the bases of PWs if we need to describe
scattering states well. The localized orbitals as well as the Wannier
functions can hardly fill the vacuum region systematically.

Here we will introduce new bases named as the contracted plane
waves (CPWs). CPWs satisfy the periodic gauge,
thus are represented by the Bloch sum of the non-orthogonalized localized bases.
CPWs are generated easily as the linear combinations of PWs.
Thus we will be easily make IinBZ.
As an another application,
CPWs can be used as bases for the first-principles calculations.
Especially, we expect CPWs can be easily included as a part of bases
in the Gaussian-based packages as {\it Crystal} \cite{crystal}.

\section{Construction of the set of CPWs}
\subsection{Definition of CPW}
\label{defcpw}
Bases in the set of CPWs $\{P_{\bfq n}(\bfr)\}$ are defined as
\begin{eqnarray}
&& P_{\bfq n}(\bfr)= \frac{1}{\sqrt{N_n}}\sum_\bfG C_n(\bfq+\bfG)
 \exp(i (\bfq+\bfG) \bfr) \label{eq:defcpw} ,\\
&& C_n(\bfq+\bfG)= \nonumber \\
&& \exp\left(- \frac{1}{2\alpha^2}
  \sum_{i,j}(q_i+G_i - \bar{G}_{nj}) A_{ij} (q_j+G_j - \bar{G}_{nj}) \right)\label{eq:cnfac},
\end{eqnarray}
where $\bfq$ is the wavevectors in the BZ, $q_i$ denotes the x,y,z components
for $\bfq$; $G_i$ and $\bar{G}_{ni}$ are for $\bfG$ and $\bar{\bfG}_n$, as well.
$\bar{\bfG}_n$ is a vector in a set $\bar{\Omega}=\{\bar{\bfG}_n|n=1,2,3,... N_{\rm max}\}$;
we show how to choose $\bar{\Omega}$ and $\alpha$ in Sec.\ref{cpwpara}.
In \req{eq:cnfac}, we take sum for all the reciprocal vectors $\bfG$.
$\sqrt{N_n}$ are normalization factors so that $\int d^3r (P_{\bfq
n}(\bfr))^2 =1$.
Since the symmetric matrix $A_{ij}$ is determined for the Bravais lattice as follows,
the set $\{P_{\bfq n}(\bfr)\}$ is specified by the parameters $\alpha$ and $\bar{\Omega}$.

$A_{ij}$ is an invariant symmetric matrix under the 
symmetry of the Bravais lattice like the dielectric constants.
While $A_{ij}$ is essentially trivial except a constant factor in the case of simple lattice,
let us consider case of general Bravais lattice.
We ask that the spheroid given by $\sum_{i,j}q_iA_{ij}q_j=1$ should give
an optimum fitting of the 1st Brillouin zone (BZ).
We have a few possible options to determine $A_{ij}$.
One is that the spheroid can be given as the largest spheroid
inside the 1st BZ, one another is that we can determine $A_{ij}$
to reproduce the 'moment of inertia' of the 1st BZ.
In either way, we can obtain three orthonormalized vectors
$\bfn_1$, $\bfn_2$, and $\bfn_3$ with corresponding eigenvalues
$c_1,c_2$, and $c_3$ for such $A_{ij}$. Thus
we can interpret that $A_{ij}$ specify an approximation of the 1st BZ by a cuboid
where as we scale the size of $A_{ij}$ so that $c_1 c_2 c_3 =|A_{ij}|=\left(\frac{2^3}{V_{\rm BZ}}\right)^2$.

We can easily see that the CPWs satisfy the periodicity of the usual Bloch functions as
\begin{eqnarray}
&& P_{\bfq n}(\bfr+ \bfT)= P_{\bfq n}(\bfr ) \exp(i \bfq \bfT), \\
&& P_{\bfq+\bfG n}(\bfr) = P_{\bfq n}(\bfr) \label{eq:Gperiodic},
\end{eqnarray}
where $\bfT$ is the translation vectors.
We can define $P_{\bfq n}(\bfr)$ even for the augmented plane
waves (APWs) if we use APWs instead of $\exp(i (\bfq+\bfG) \bfr)$.
It is virtually possible to take infinite sum for all $\bfG$ numerically,
because of the truncation due to the Gaussian factor in \req{eq:cnfac}.
\req{eq:cnfac} shows that a PW whose $\bfq+\bfG$ is nearest to $\bar{\bfG}_n$ has largest $C_n(\bfq+\bfG)$.
Thus we may say that $P_{\bfq n}(\bfr)$ is similar with
$\exp(i \bar{\bfG}_n \bfr)$ when $\alpha$ is small enough.
As we noted in Sec.\ref{intro}, one of the usage of CPWs is for IinBZ.
Because $P_{\bfq n}(\bfr)$ satisfies the periodic gauge condition \req{eq:Gperiodic},
quantities such as $\langle P_{\bfq n} | H | P_{\bfq n'} \rangle$,
is periodic for $\bfq$ in the BZ. This allows us to make IinBZ.

From \req{eq:defcpw}, we have real-space localized functions
$P_{n}(\bfr)$ as
\begin{eqnarray}
P_{n}(\bfr)&=& \sum_\bfq P_{\bfq n}(\bfr) 
= \frac{1}{\sqrt{N_n}} (2\pi\alpha)^{3/2}  \nonumber \\
&&\times \exp (-  \frac{\alpha^2 \sum_{i,j}r_j A^{-1}_{ij} r_j}{2} + i \bar{\bfG}_n \bfr ).
\label{eq:local}
\end{eqnarray}
When we use real $\alpha$, the center of $P_{n}(\bfr)$ is located at $\bfr=0$.
Note that bases in the set $\{P_{n}(\bfr) \}$ are not orthogonalized. Thus we need
overlap matrix $O_\bfq(n,n')=\langle P_{\bfq n}(\bfr)| P_{\bfq n'}(\bfr)
\rangle$ to handle the set.
CPWs are nothing but the Bloch sum of the Gaussians with oscillations as
shown in \req{eq:local}.

To use CPWs in the first-principles calculations, especially for IinBZ,
we have following requirements for the set $\{P_{\bfq n}(\bfr)\}$.
\begin{enumerate}
 \item[(1)] Each $P_{\bfq n}(\bfr)$ should be smoothly changing as a
	    function of $\bfq$ in the BZ.
 \item[(2)] The low-energy Hilbert space spanned by
	    $\{ \exp(i (\bfq+\bfG) \bfr)| \bfG \in \Omega_{\rm S}(\bfq) \}$ should
	    be contained well in the Hilbert space spanned by $\{P_{\bfq n}(\bfr)\}$,
            where $\Omega_{\rm S}(\bfq)$ is a set of $\bfG$ near $\bfq+\bfG=0$.
 \item[(3)] Linear-dependency of bases are kept. That is, the
	    eigenvalues of overlap matrix $O_\bfq(n,n')$ should be not
	    too small for double precision calculations.
\end{enumerate}
Under the assumption that $\{ \exp(i (\bfq+\bfG) \bfr)| \bfG \in \Omega_{\rm S}(\bfq) \}$ is 
good enough to expand eigenfunctions, the condition (2) assures that
the space spanned by $\{P_{\bfq n}(\bfr)\}$ is good enough, too.
In fact, we expect not so large $\Omega_{\rm S}$ is required 
to obtain accurate bands in $GW$ calculations \cite{kotani_formulation_2015}. 
Together with the condition (1) and (3), we can use $P_{\bfq n}(\bfr)$ as bases to
expand the one-body Hamiltonian for the interpolation.

Although our Hilbert space spanned by the set $\{P_{\bfq n}(\bfr)\}$ is
not satisfying the translational symmetry,
we will see that the translational symmetry can be recovered well as shown in Sec.\ref{test}.

\subsection{Parameters to specify a set of CPWs}
\label{cpwpara}
To specify the set $\{P_{\bfq n}(\bfr)\}$, we need $\bar{\Omega}$.
$\bar{\bfG}_n$ in $\bar{\Omega}$ is given by
\begin{eqnarray}
&&\bar{\bfG}_n   =  \frac{1}{\beta}  \bfG(n),        \label{eq:scaling}
\end{eqnarray}
where $\bfG(n)$ is the reciprocal $\bfG$ vectors satisfying
$|\bfG(n)|<G_{\rm MAX}$. Thus the number of bases in the set
$\{P_{\bfq n}(\bfr)\}$ is given by $G_{\rm MAX}$.
$\beta$ is a scaling factor, a little larger than unity.
The value of $\beta$ is examined in Sec.\ref{test}.
This procedure of \req{eq:scaling} gives a little denser mesh of
$\bar{\bfG}_n$ than the mesh of $\bfG$.
As we see in Sec.\ref{test}, the scaling procedure \req{eq:scaling}
is essential to reproduce eigenfunctions around the BZ boundaries accurately.

In addition, we need to determine the parameter $\alpha$. The condition (1) in
Sec.\ref{defcpw} requires that $\alpha$ is large enough so that
$C_n(\bfq+\bfG)$ is a smooth function of $\bfq$, while the condition (2) requires
$\alpha$ is small enough so that $C_n(\bfq+\bfG)$ chooses one of $\bfq+\bfG$ dominantly.
To let these requirements balanced, we use a balancing condition
between the damping factor in \req{eq:cnfac} measured by the unit of the $\bfG$-lattice spacing,
and the damping factor of \req{eq:local} measured by the realspace-lattice spacing. 
Because we have approximated the 1st BZ by the cuboids given
by $A_{ij}$ ( real-space cuboid by $A^{-1}_{ij}$ as well), we have
\begin{eqnarray}
 \frac{-4}{2 \alpha^2} \approx  \frac{-\alpha^2 (\pi)^2}{2}  . \label{eq:grref}
\end{eqnarray}
Thus we have $\alpha = \sqrt{2/\pi} \approx 0.8$.
We use this value in the test calculations shown in Sec.\ref{test}.


With the three parameters $G_{\rm MAX},\alpha$ and $\beta$,
We can give the set $\{P_{\bfq n}(\bfr)\}$
for given primitive cell vectors of a crystal structure.
In Sec.\ref{test}, we will evaluate quality of the set
when we are changing these parameters.

Let us summarize the algorism to specify $\{P_{\bfq n}(\bfr)\}$.
For given $G_{\rm MAX}$, we first make a set
$\{\bfG(n)|n=1,...N_{\rm MAX}; |\bfG(n)|<G_{\rm MAX}\}$.
Then, we have $\bar{\Omega}$ by \req{eq:scaling}.
Then  \req{eq:defcpw} and \req{eq:cnfac} yields $\{P_{\bfq n}(\bfr)\}$, whereas
$A_{ij}$ is given in advance so that the 1st BZ is approximated by
the cuboid specified by $A_{ij}$.
The number of $\bfG$ for the sum in \req{eq:cnfac}
should be large enough so that the sum converges well.

There is a possibility to use another algorism to determine
$\bar{\Omega}$; furthermore $\alpha$ can be $n$-dependent.
In fact, we have tested some other possibilities such as 
$\bar{\bfG}_n  = \frac{|\bfG(n)|}{|\bfG(n)| +\beta}  \bfG(n)$ instead of
\req{eq:scaling}, however we did not observed meaningful improvements in
the tests shown in Sec.\ref{test}. Thus we focus on the
algorism presented here.


\section{Numerical Tests and Determination of parameters of CPWs}
\label{test}
Here we perform test calculation for the fcc empty lattice.
In the followings, we consider a case of the fcc lattice for Si.
The size of primitive cell is $\frac{a^3}{4}$ where $a=5.43$\AA.
We show that the set $\{P_{\bfq n}\}$
spans the low energy part of the Hilbert space spanned by PWs very well,
as long as we take adequate choice of $(\alpha,\beta)$.

\begin{figure}[ht]
 \caption{
 We plot $\epsilon^m_\bfq$ in \req{eq:empty} for the fcc empty lattice, that is,
 the band structure of empty lattice calculated with the set of CPWs $\{P_{\bfq n}\}$.
 We plot the exact empty bands
 $\epsilon^{m,{\rm Exact}}_\bfq =\frac{\hbar^2(\bfq+\bfG)^2}{2m}$ with purple solid lines together.
 At low energy, we see that $\epsilon^m_\bfq$ clearly on the solid
 lines. Three panels are for $\beta=$1.0,1.6 and 2.0,
 while $G_{\rm MAX}=$4.0 Ry and $\alpha=$0.8.}
 \includegraphics[width=9cm]{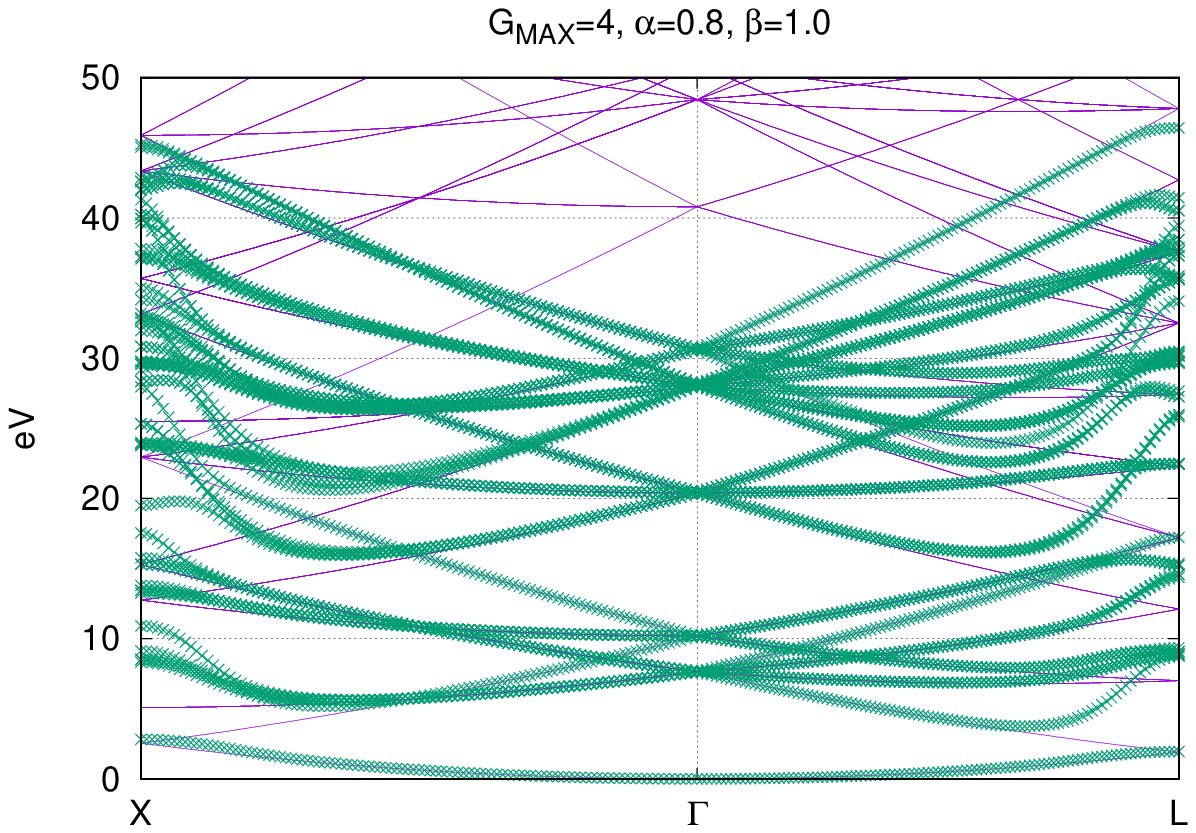}
 \includegraphics[width=9cm]{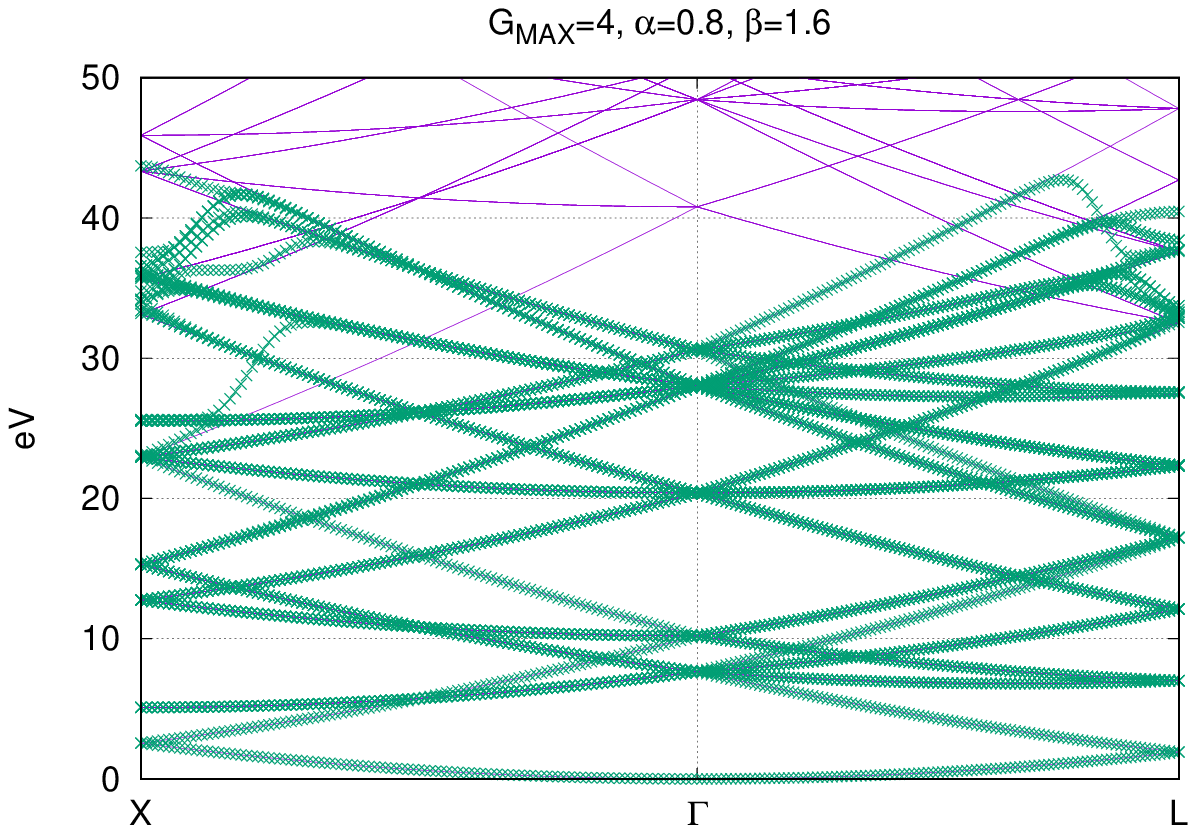}
 \includegraphics[width=9cm]{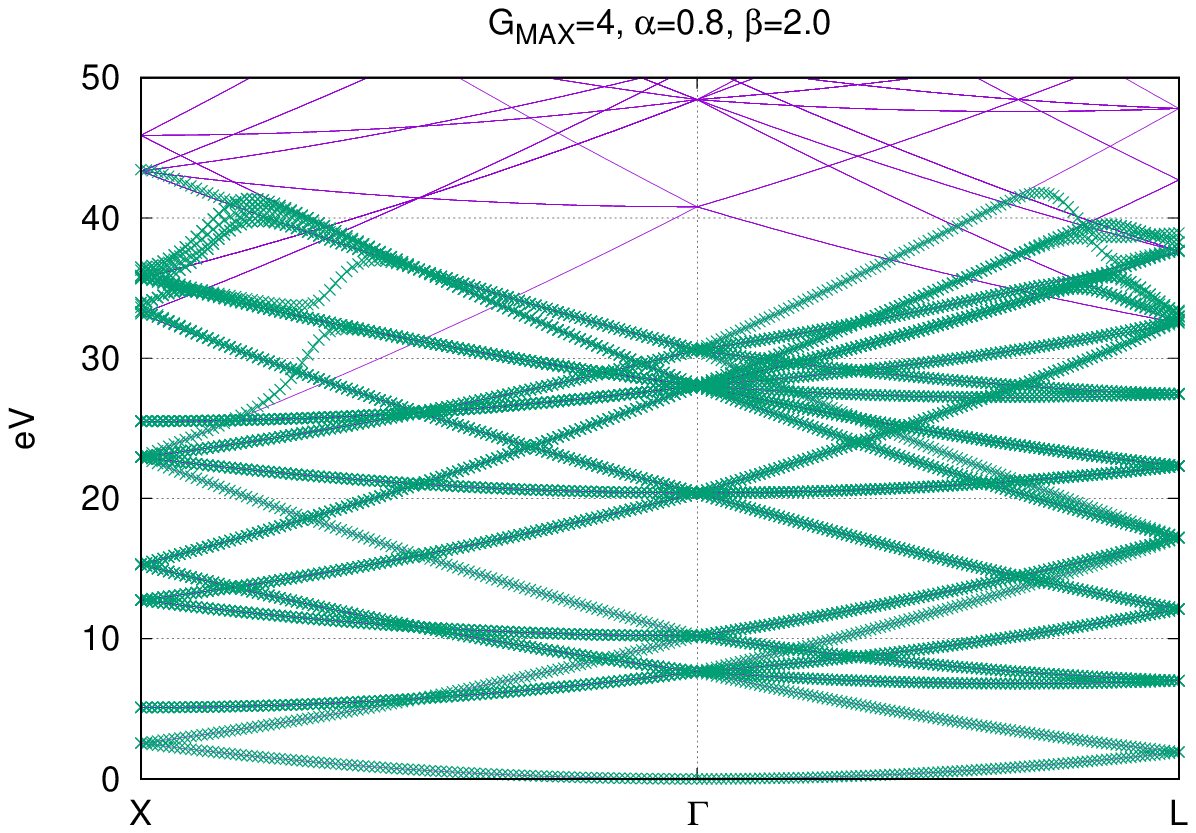}
\label{fig:bandsc}
\end{figure}

We first look into eigenvalues.
We solve the following eigenvalue problem to determine eigenvalue $\epsilon^m_\bfq$ for empty lattice;
\begin{eqnarray}
 \sum_{n'} \Large( \langle P_{\bfq n}| \frac{-\nabla^2}{2m} | P_{\bfq n'} \rangle 
 - \epsilon^m_\bfq
 \langle P_{\bfq n}| P_{\bfq n'} \rangle \Large) a^m_{\bfq n'} = 0. \label{eq:empty}
\end{eqnarray}
In Fig.\ref{fig:bandsc}, we plot $\epsilon^m_\bfq$ for the fcc empty lattice
with $\alpha=0.8$ and with $\beta=1, 1.6, 2.0$.
We use $G_{\rm MAX}=4$, which determines the number of bases is 56.
We show exact eigenvalues of the empty lattice by solid lines together.

We see disagreements at the BZ boundaries for the case of no
scaling, $\beta=1$. However, we see 
good agreement with the exact ones for $\beta=1.6$ and $\beta=2.0$ in the whole BZ.
We show details of agreement afterwards in Fig.\ref{fig:cpwG}.
In the middle panel of $\beta=1.6$, we see a branch from $\Gamma$ to
X are bending at $~$32 eV and getting to be $~23$eV at X point.
This is reasonable because of \req{eq:cnfac};
when we are changing $\bfq$, $\bfG$ giving the largest $|\bfq+\bfG-\bar{\bfG}_n|$ switches.
For larger $\beta$, we have better agreement at low energies,
but we have more bumpy bands at higher energies.
Since larger $\beta$ gives denser bases for lower energy, larger $\beta$ shows
better agreement at low energy but worse at high energy.

\begin{figure}[ht]
 \caption{
 Similar plots in Fig.\ref{fig:bandsc}. Three panels are for $\alpha=$0.6, 0.8, and 1.0,
 while $G_{\rm MAX}=$4.0 Ry and $\alpha=$0.8. Middle panel is the same
 as that in Fig.\ref{fig:bandsc}.}
 \includegraphics[width=9cm]{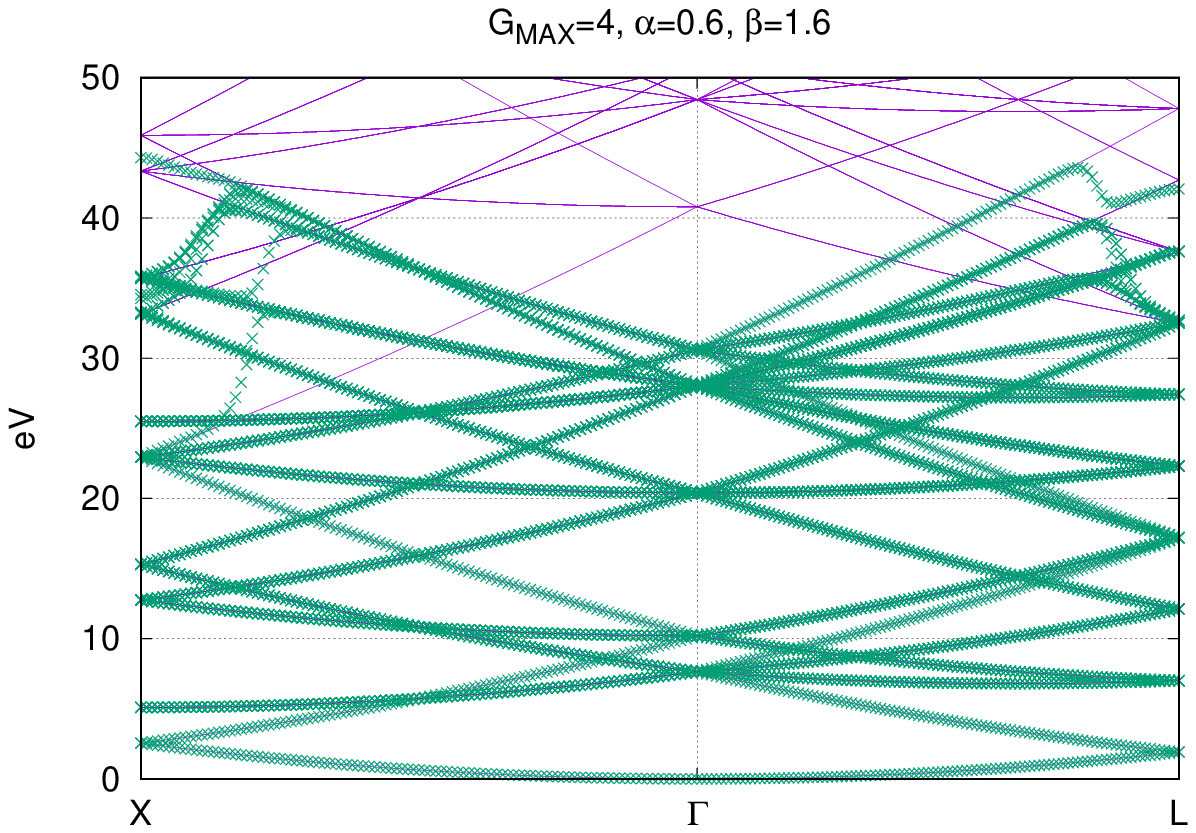}
 \includegraphics[width=9cm]{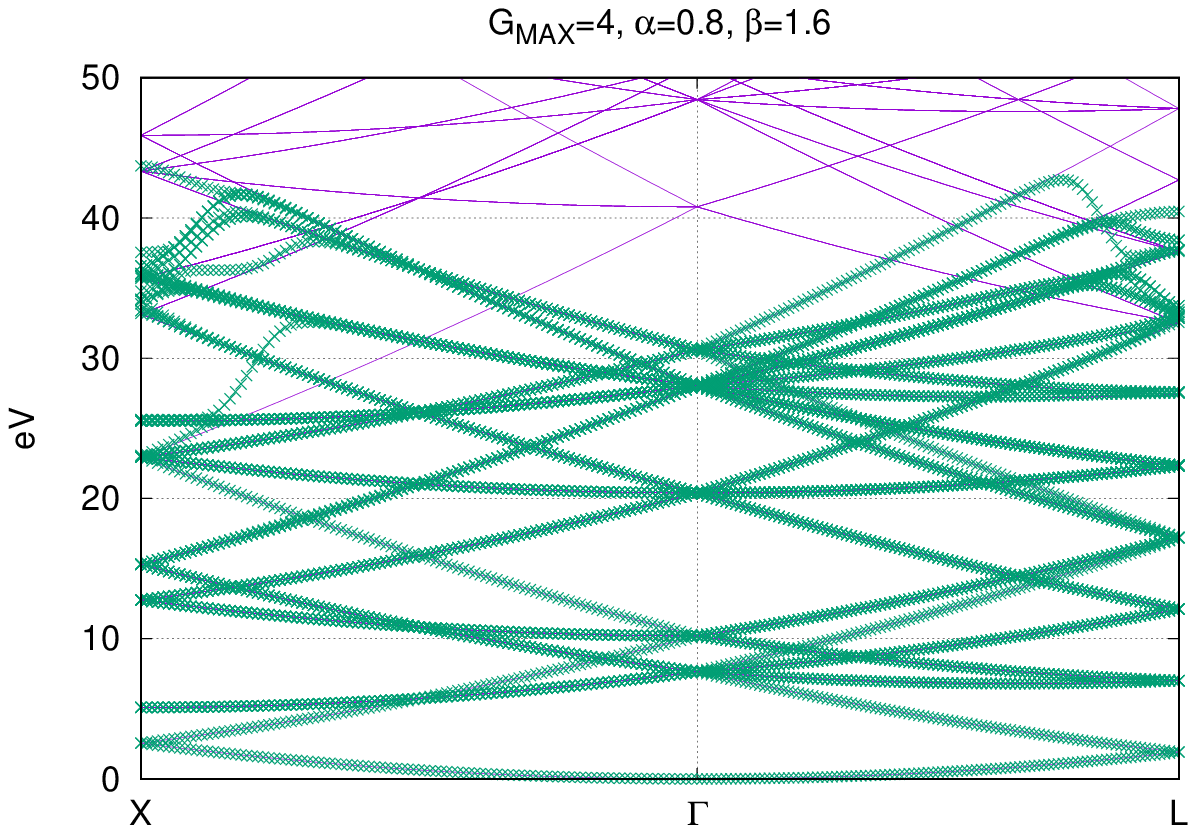}
 \includegraphics[width=9cm]{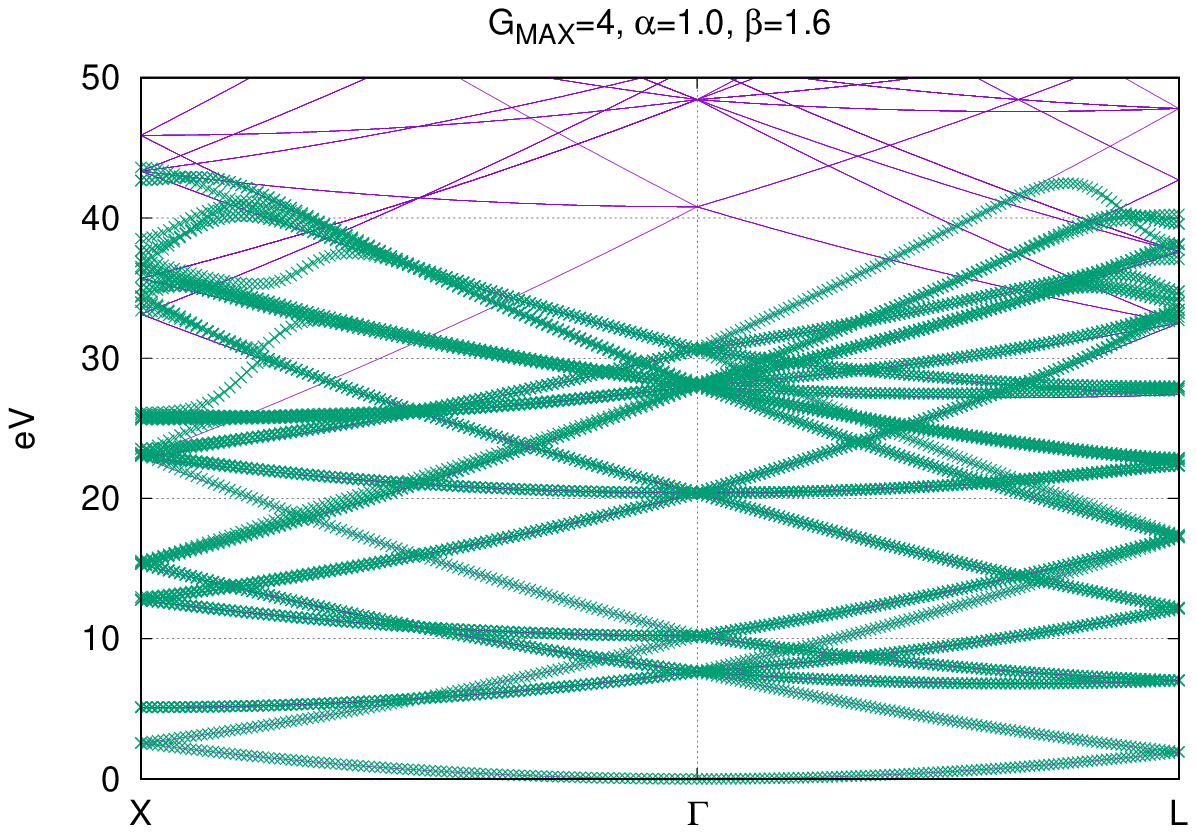}
\label{fig:bandgr}
\end{figure}

Let us examine the effect of parameter $\alpha$.
In Fig.\ref{fig:bandgr}, we show how the bands change for changing $\alpha$.
The middle panel of Fig.\ref{fig:bandgr} is the same as that in Fig.\ref{fig:bandsc}.
As we expect from \req{eq:cnfac}, we have better smoothness of energy
bands in the BZ for larger $\alpha$.

\begin{figure}[ht]
 \caption{Reproductivity of PWs. For $\frac{\hbar^2(\bfq+\bfG)^2}{2m}$
 (y-axis) along X-$\Gamma$-L, we plot the projection weight
 $Q(\bfq+\bfG)$. Lines for all branches are plotted.
 In this case, $Q(\bfq+\bfG)$ below $\sim$ 30 eV are almost one-hundred percent.}
 \includegraphics[width=10cm]{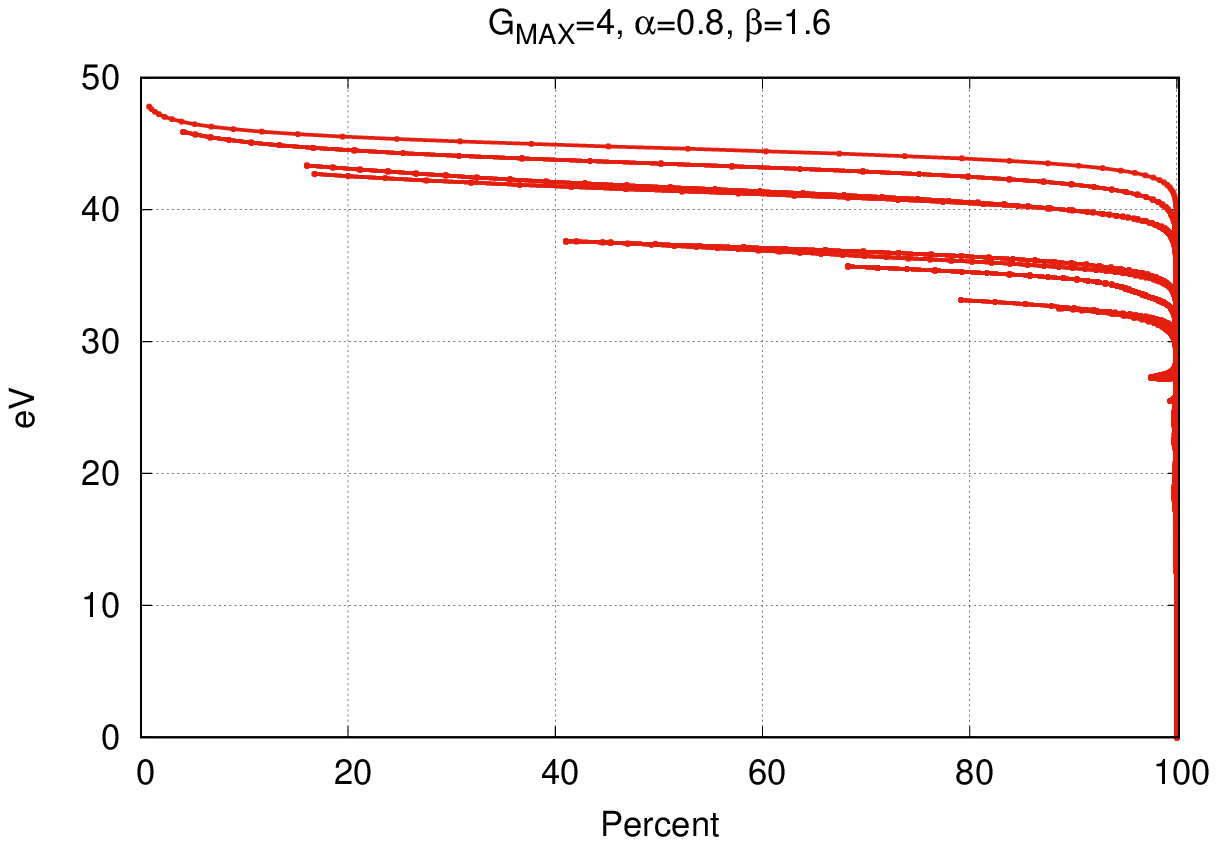}
\label{fig:percent}
\end{figure}

Another check is the ability to reproduce PWs as superpositions of $\{P_{\bfq n}\}$.
For this purpose, we calculate the square of the amplitude of projection to the
space of PWs, $Q(\bfq+\bfG)= \sum_n |\langle P_{\bfq n} |\exp(i (\bfq +\bfG) \bfr) \rangle|^2$.
bWe show a plot, eigenvalue $\frac{\hbar^2(\bfq+\bfG)^2}{2m}$ (y-axis)
vs. $Q(\bfq+\bfG)$ (x-axis), in Fig.\ref{fig:percent}.
Lines from multiple branches of energy bands are plotted.
This shows most of all $Q(\bfq+\bfG)$ below $\sim 30$ eV is almost at one-hundred percent.
This means that such low energy part is well-expanded by $\{P_{\bfq n}\}$.

\subsection{Determination of parameters}
Based on the condition in Sec.\ref{defcpw} and observations in Sec.\ref{test},
we discuss how to determine parameters optimally.
For given $G_{\rm MAX}$, we have to determine $(\alpha,\beta)$.

\begin{figure}[ht]
 \caption{Maximum error of eigenvalues and minimum value of
 overlap matrix for varieties of $(\alpha,\beta)$. 
 The top panel is for $G_{\rm max}=4$ which gives 58 number
 of bases as shown in its title. Other panels as well.
 Horizontal and Vertical red lines are guide for
 eye; right-bottom area suggests preferable $(\alpha,\beta)$.}
 \includegraphics[width=9cm]{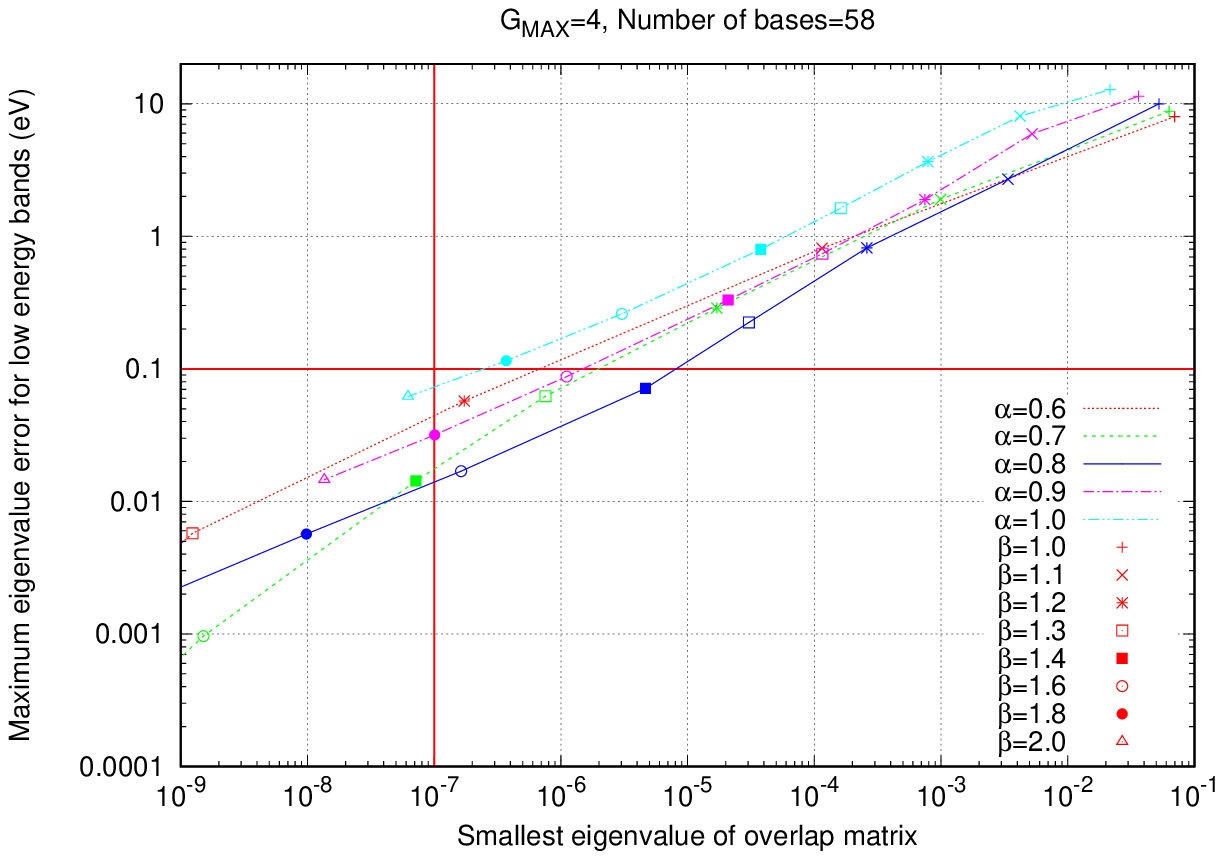}
 \includegraphics[width=9cm]{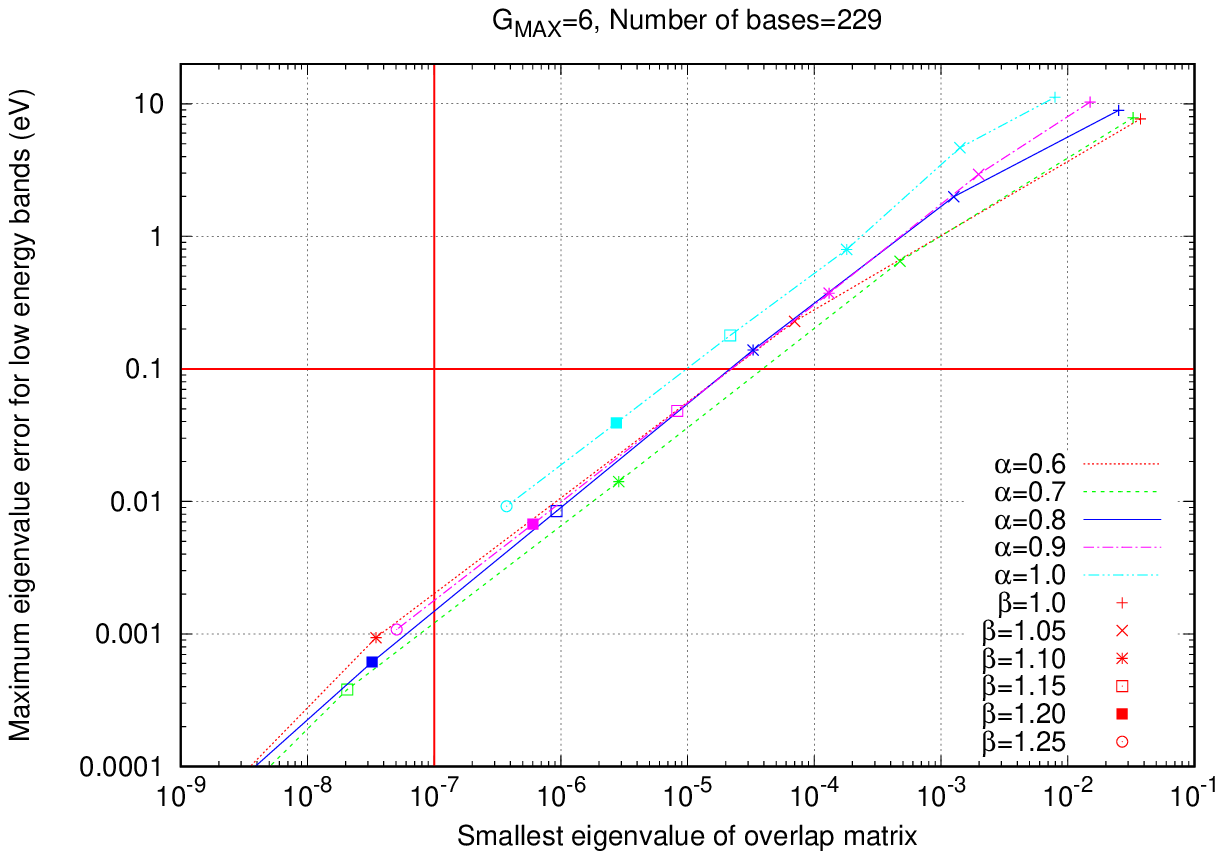}
 \includegraphics[width=9cm]{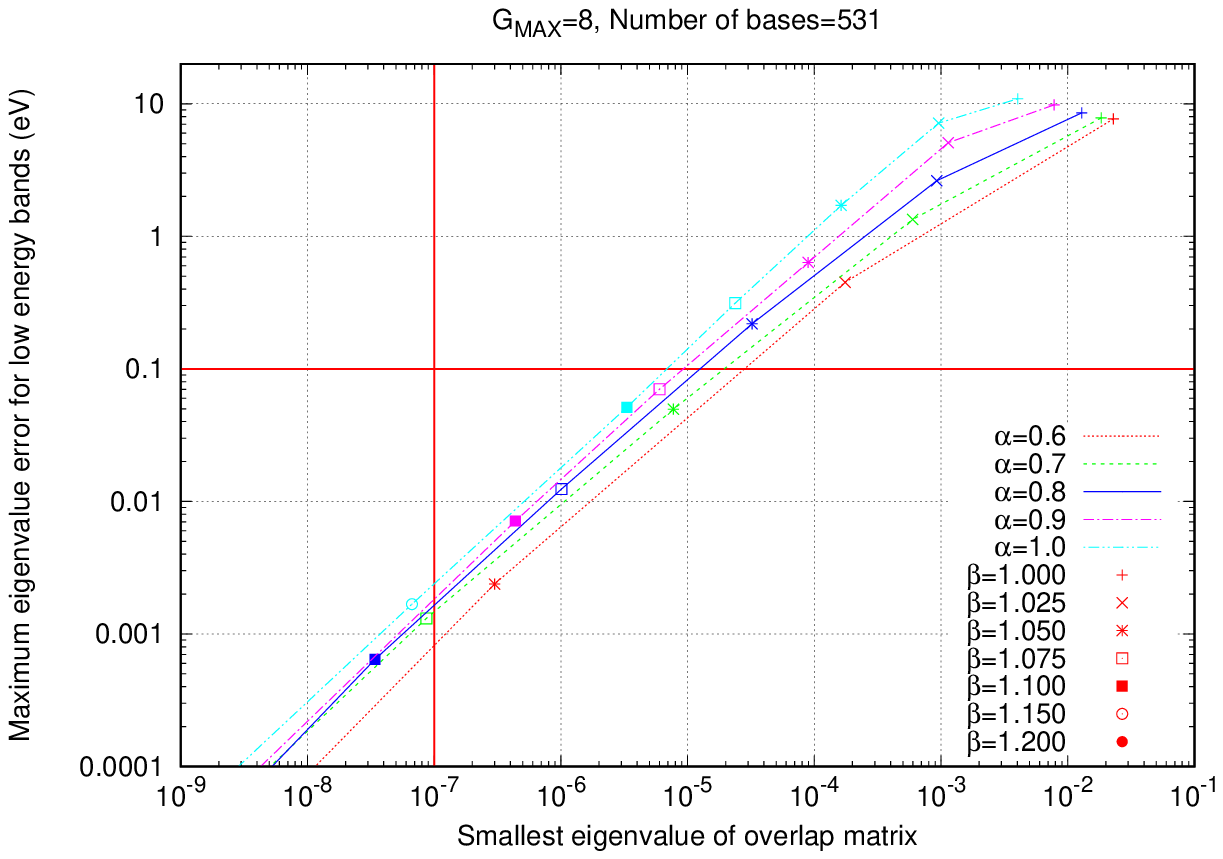}
\label{fig:cpwG}
\end{figure}

For this purpose, we plot maximum error of eigenvalues at low energies
(eigenvalues below 20eV) and the smallest size of the eigenvalues of the overlap
matrix in Fig.\ref{fig:cpwG} for given $(\alpha,\beta)$. 
Each line is for each $\alpha$. Along the line, we have different symbols for different $\beta$.
The top panel is for $G_{\rm MAX}=4$. For example, we can see that 
the error is $\sim$0.02eV for $(\alpha=0.8,\beta=1.6)$.
Fig.\ref{fig:cpwG} shows that enlarging $\beta$ can efficiently
reduce error at low energy. 
However, because of round-off error in numerical calculations,
it is safer to use $(\alpha,\beta)$ which gives not too small eigenvalues of
the overlap matrix.

Thus optimum $(\alpha,\beta)$ may be chosen so that it is at right-bottom in
the panels. The discussion around \req{eq:grref} suggested $\alpha=0.8$.
We can see that behavior of results shown in Fig.\ref{fig:cpwG} looks stable enough around $\alpha=0.8$.
Thus we claim that $\alpha=0.8$ is not a bad choice.
Then we determine $\beta$ showing not too small eigenvalue of the overlap matrix.
If we set the smallest eigenvalue is $10^{-7}$ (on the red vertical line),
we can see $\beta \sim 1.07$ from the line of $\alpha=0.8$ in the case of bottom panel $G_{\rm MAX}=8$.
Thus we suggest a prescription to determine $\alpha$ and $\beta$ for
given $G_{\rm MAX}$; use $\alpha=0.8$, and determined $\beta$ so that the
smallest eigenvalue is not too small. This determination can be done by test
calculations or a table prepared in advance.
We think this can be easily implemented.

\subsection{CPWs in the supercell}
Although the procedure to give a set of CPWs do not depends on the cell size,
we have one another choice to give the set of CPW for a supercell in the
following manner, based on the fact that the supercell is made from small cells.
The primitive vectors of the supercell $\bfA_i$ are given as $\bfA_i= \sum_j n_{ij} \bfa_j$
by the $3\times 3$-integer matrix $n_{ij}$,
where $\bfa_j$ is the primitive vectors of a small cell such as the fcc lattice.
The volume of the supercell cell is $N_{\rm S}=|n_{ij}|$ times
larger than that of the small cell.

To give a set of CPW of a supercell, we first generate
the localized functions \req{eq:local} for the small cell.
Then we put the localized functions at the origins of the small cells.
Thus the number of the localized functions contained in the supercell is
$N_{\rm S} \times N_{\rm MAX}$, where $N_{\rm MAX}$ is the number of
bases for the small cell.
Then we can use the back Fourier-transformation to obtain the representation 
\req{eq:defcpw}. This back transformation is done
for $\bfq$ and $\bfG$ for the supercell as well.
Thus we ends up with the CPWs for the supercell as $\{P_{\bfq n i}\}$,
where $i=1,2,...,N_{\rm S}$ and $n=1,2,...,N_{\rm MAX}$.
By definition of this construction, the set of CPWs for the supercell
exactly reproduces the result for the set of CPWs for the small cell.
An example is the case of antiferro-II NiO, where we have
$N_{\rm S}=2$ for small cell of the fcc lattice.
It will be possible to decompose small and almost-isotropic cells.

\section{discussion and summary}

The set of CPWs will be very useful to make interpolation in the BZ.
For the matrix as a function of $\bfq$ expanded in PWs/APWs, we can
easily re-expand them in the set of CPWs
since the transformation matrix between PWs and CPWs is explicitly given.
Then the interpolation becomes easy because of the periodicity of CPWs
in the BZ. This procedure via CPWs allows
automatic interpolation which was difficult in the Wannier interpolation.

Since CPWs are the oscillating Gaussians in real space as shown in
\req{eq:local}, CPWs can be relatively
easily used in the Gaussian-based methods such as {\it Crystal} \cite{crystal}.
With CPWs, we can obtain high-energy states easily
without being bothered with the choices of the Gaussian basis sets.
The idea of CPWs might be a key to bridge the Gaussian bases and the PW bases.
In addition, CPWs can directly used in the usual first-principles methods
such as PP, LAPW, and PP methods.
The real-space representation of CPWs may work for reducing the number of the bases
when we treat a supercell with large vacuum region.

We did not yet get aware serious problems to implement CPWs to practical methods.
Although the set of CPWs does not satisfy the translational symmetry
completely, it will cause little problem since the low-energy part of
the Hilbert space spanned by the PWs are well-reproduced. 
Determination of parameters to specify a set of CPWs is not difficult,
thus it can be automatic without tuning by human.


\begin{acknowledgments}
This is supported by JSPS KAKENHI Grant Number 17K05499.
T.Kotani thanks to discussion with Prof. Hirofumi Sakakibara.
We also thank the computing time provided by Research Institute
for Information Technology (Kyushu University).
\end{acknowledgments}

\bibliography{refkotani}

\end{document}